\documentclass[prd,groupaddress,nofootinbib]{revtex4}
\usepackage{framed}
\usepackage{bbold}
\usepackage{slashed}
\usepackage{graphicx}
\usepackage{amsmath}
\usepackage{amssymb}
\usepackage{epsfig}
\usepackage{hhline}
\usepackage{soul}
\usepackage{tabularx,pifont,multirow}

\newcommand{\be}{\begin{equation}}
\newcommand{\ee}{\end{equation}}

\begin{document}

\preprint[\leftline{KCL-PH-TH/2018-{\bf 67}}

%

\title{\Large {\bf Finite-energy dressed string-inspired Dirac-like monopoles} }

\bigskip

\author{Nick E. Mavromatos and Sarben Sarkar}

\vspace{0.5cm} 

\affiliation{Theoretical Particle Physics and Cosmology Group, Department of Physics, King's College London, Strand, London WC2R 2LS, UK}


\begin{abstract}
\vspace{0.5cm}
\centerline{\bf Abstract }

On extending the Standard Model (SM)  Lagrangian, through a non-linear Born-Infeld (BI) hypercharge term with a parameter $\beta$ (of dimensions of [mass]$^2$),  a \emph{finite} energy monopole solution was claimed by Arunasalam and Kobakhidze~\cite{AK} . We report on a new class of solutions within this framework which was missed in the earlier analysis. This new class was discovered on performing consistent analytic asymptotic analyses of the nonlinear differential equations describing the model; the shooting method used in numerical solutions to boundary value problems for ordinary differential equations is replaced in our approach by a method which uses  diagonal Pad\'e approximants. Our work uses the ansatz proposed by Cho and Maison~\cite{CM} to generate a static and spherically symmetric monopole with finite energy and differs from that  used in the solution of \cite{AK}. Estimates of the total energy of the monopole are given, and detection prospects at colliders are briefly discussed. 

\end{abstract}
\maketitle

\section{Introduction \label{sec:intro}}
It is a curious fact that Maxwell initially wrote as one of his eponymous equations:
\begin{equation}
\label{fe1}
curl\,\vec A = \mu \vec H.
\end{equation}
The magnetic field is denoted by $\vec H$; $\vec A$ is the electromagnetic vector potential and $\mu$ is the magnetic permeability.    If $\vec A$ is a non-singular vector then 
\begin{equation}
\label{fe2}
\vec \nabla \cdot \vec H = 0.
\end{equation}
In 1931 Dirac~\cite{dirac} considered a singular $\vec A$ field. If $\vec n$ is a constant unit vector in the $z$-direction,
then the vector potential has the form
\begin{equation}
\label{fe3}
\vec A\left( {\vec r} \right) = g^m\, \frac{{\vec r \times \vec n}}{{r\left( {r - \vec r \cdot \vec n} \right)}},
\end{equation}
with $g^m$ the magnetic charge~\cite{dirac}, 
which has a singularity along the positive $z$-axis known as a Dirac string.\footnote{It is also useful to consider the expression for $\vec A$ in spherical polars (with the co-ordinates ${r,\theta,\phi}$): 
\[\vec A\left( {\vec r} \right) = \left( {1 + \cos \theta } \right)\frac{i}{e}{U^{ - 1}}\vec \nabla U\] where
$U = \exp \left( { - ieg^m\, \phi } \right)$. $U$ is a singular gauge transformation. } On calculating 
$curl\,\vec A$ we find a magnetic field of a monopole
\begin{equation}
\label{fe4}
\vec B\left( {\vec r} \right) = g^m \, \tfrac{{\vec r}}{{{r^3}}}
\end{equation}
except along the string. In classical physics the string should play no role in the dynamics of particles because of its infinitesimal nature. In quantum theory the situation is different since the wave function $\Psi $ of a non-relativistic particle with charge $e$ and mass $m$ in a monopole background satisfies
\begin{equation}
\label{fe5}
i\frac{\partial }{{\partial t}}\Psi  =  - \frac{{{{\left( {\vec \nabla  - ie\vec A} \right)}^2}}}{{2m}}\Psi 
\end{equation}
which has the stationary solution
\begin{equation}
\label{fe6}
\Psi \left( {\vec r} \right) = {\Psi _0}\left( {\vec r} \right)\exp \left[ {ie\int\limits_0^{\vec r} {d\vec x.\vec A\left( {\vec x} \right)} } \right].
\end{equation}
The wave function $\Psi \left( {\vec r} \right)$ is single valued and so there should be no change in $\Psi \left( {\vec r} \right)$ for  a small circuit $C$ around the string. The change in the argument of the exponential in (\ref{fe6}) along $C$  should be of the form $2n\pi i$ where $n$ is an integer. This leads to the condition 
\begin{equation}
\label{fe7}
e\oint_{C} {d\vec x \cdot\vec A\left( {\vec x} \right) = 2\pi n}.
\end{equation}
Since the magnetic flux through the circuit $C$ is $4 \pi g^m$, from (\ref{fe7}) we deduce deduce the celebrated Dirac charge quantisation condition~\cite{dirac}
\begin{equation}
\label{fe8}
eg^m= \frac{n}{2}, \quad n \in {\mathbf Z} \, {\rm U} \, \{0\},
\end{equation}
or, equivalently 
\begin{equation}\label{dqc}
g^m= \frac{1}{2\,\alpha} \, n \, e = g_D n, \quad g_D \equiv \frac{1}{2\,\alpha}, \quad n \in {\mathbf Z} \, {\rm U} \, \{0\},
\end{equation}
where $\alpha $ is the fine structure constant, which at low (strictly speaking zero) energy has the value $\alpha=1/137$, and 
$g_D = 68.5$ is the fundamental Dirac charge.
 
The Standard Model of particle physics has been developed subsequent to the original work of Dirac. It describes the weak, electromagnetic and strong interactions of leptons and hadrons. It is important to ascertain how monopoles may fit into the electro-weak sector of the SM, in part because of the possible detectability of electro-weak monopoles (of mass a few TeV) at colliders in the near future. The work of `t Hooft and Polyakov~\cite{hpmono} provides a detailed paradigm on magnetic monopole soliton solutions, which arise in quantum field theories with simple gauge groups (such as SU(3) and Grand Unified groups SU(5)), under spontaneous symmetry breaking. In such solutions, Dirac's quantisation arises as a topological property of mappings associated with the solution and not because of a Dirac string. 

The SM does not have a simple group, as the gauge group of the electroweak sector is  $SU(2) \times U_Y(1)$, with $U_Y(1)$ the weak hypercharge gauge group. This was one of the arguments against any attempts to find a topological monopole solution within the SM. 
Nevertheless, Cho and Maison (CM)~\cite{CM} presented a monopole solution within the $SU(2) \times U_Y(1)$ electroweak theory, by arguing that  a non-trivial topology of th solutions was still possible due to an underlying $CP^1$ structure. Specifically,  
 Cho and Maison view the normalised Higgs doublet field of the SM in the symmetry broken phase as a $CP^1$ field, which is known to have a non-trivial second homotopy $\Pi_2(CP^1)={\tt \mathbb{Z}}$; it was argued~\cite{CM} that this homotopy would lead to a charge quantisation of the monopole as in the 'Hooft-Polyakov case.
 In \cite{CM} the monopole and dyon solutions (characterised by both electric and magnetic charges), suffer, however, from ultraviolet infinities in their total energy. This casts serious doubts on the existence of consistent soliton solutions (which by definition have finite energy).

The considerations above suggest that the model needs a modification before any physical conclusions can be drawn drawn. Examples of a modification of the theory beyond the SM, are inclusion of non-minimal couplings of the Higgs field with the hypercharge kinetic terms in the effective Lagrangian~\cite{CKY,you}, 
or through higher derivative extensions, such as  
a non-linear Born-Infeld gauge field theory, which notably arises as a low energy field theory limit of strings~\cite{gsw}.  
Although in string theory, the full standard model gauge group admits such non-linear extensions, nonetheless, it suffices for our phenomenological purposes to restrict our attention only to the Born-Infeld extension of the hypercharge sector, and seek monopole solutions of CM type, following ~\cite{AK}~\footnote{Prior to the work of \cite{AK}, monopole solutions within the framework of Born-Infeld electrodynamics , but different from the CM case, have been discussed in \cite{initialBI}.} . 

The finiteness of the monopole solution in  Born-Infeld type theories is an immediate consequence of the finiteness of the electromagnetic field 
energy density in Born-Infeld non-linear electrodynamics~\cite{zwiebach}.The identification of finite energy consistent monopole solutions in (extensions of) the SM, represents not only an important theoretical advance, but also an important step  towards a consistent phenomenology since one can provide estimates of the total energy/mass of monopoles and thus check the feasibility of their production at colliders~\cite{vento}.
Recently experimental efforts to discover monopoles have redoubled~\cite{atlas,moedal}. In particular, searches for magnetic monopoles of lowest magnetic charge are ongoing in the ATLAS-LHC experiment~\cite{atlas}. In addition, the MoEDAL experiment at the LHC~\cite{moedal}  is geared to the detection of highly ionising particles, among which magnetic monopoles, using a variety of experimental techniques, which allow for monopoles of high magnetic charge to be searched for for the first time in experiment.  From the arguments of Dirac the monopole magnetic charge would be an integer multiple of ${g_D}\left( { = 68.5} \right)$ (\ref{dqc}). Consequently magnetic monopoles interact \emph{strongly} with photons and are highly ionising, making TeV mass monopoles  candidates suitable for detection at MoEDAL~\cite{moedal}.  However,  as argued in ~\cite{drukier}, 
structured monopoles, such as the ones mentioned above (which arise as consistent solutions of specific quantum gauge field theories with spontaneous symmetry breaking), might exhibit extremely suppressed production cross-sections. This suppression would eliminate any prospect for detection for structured monopoles. Dirac (structureless) monopoles do not suffer from suppressed production cross-sections.  Nonetheless, finite-energy structured monopoles, might be produced abundantly from the vacuum via a thermal version of the Schwinger pair production, as advocated in \cite{rajantie}, and, hence, can still be of great relevance to heavy ion collisions at LHC. If they have sufficiently low mass the monopoles can be potentially detected by the deployment of magnetic monopole trapping detectors (of the type used in MoEDAL~\cite{moedal}) in such environments~\cite{rajantie2}. 

We now remark that the monopole solutions discussed in \cite{AK}, and also in the previous literature of the electroweak monopole~\cite{CM,CKY}  (and its finite energy extensions~\cite{you}) are based on matching numerical solutions using shooting methods. Such solutions, however, appear not to be in agreement with next to leading order analytical solutions near the monopole centre. 
It is the purpose of this work, to discuss a new class of solutions, which match an analytic asymptotic behaviour at both small and large distances from the monopole centre. The solutions are slightly different near the monopole centre from the standard electroweak monopole solutions appeared in the literature~\cite{CM,CKY,AK,you}. Nevertheless, as we demonstrate in the present article, the order of magnitude of the associated total energy (of crucial importance for their phenomenology) remains the same as in the standard CM-like case~\cite{AK}. Current experimental lower bounds of the Born-Infeld mass parameter then, imply monopole masses of at least 11 TeV, which makes such solutions relevant for potential detection only at future colliders~\cite{you2}. 

The structure of the article is as follows: in the next section \ref{sec:ewm} we introduce the model and give the dynamical equations that will be associated with monopole solutions (but not dyon solutions). In the following section, \ref{sec:dressed} we discuss our new solutions, which have not appeared before in the literature. We discuss analytic forms of the solutions for short and large distances from the monopole centre, and the associated interpolating functions, found by using Pad\'e approximant methods~\cite{AAMC,GB}. Energy estimates and thus detection prospects, are discussed in section \ref{sec:energy}. Finally, our conclusions and outlook are presented in section \ref{sec:concl}.

\section{Born-Infeld Electroweak Monopole: the set up \label{sec:ewm}}

A string-inspired  extension (ESM)  of the SM, considered in \cite{AK} and used in the current work, arises when the standard kinetic energy of the hypercharge gauge field is replaced by a \emph{non-linear }Born-Infeld term~\cite{zwiebach}.  The resultant Lagrangian is
\begin{equation}\label{SMBIL}
\mathcal{L}_{EW} = -(D_{\mu}H)^{\dagger}(D^{\mu}H)-\frac{\lambda}{2}\Big(H^{\dagger}H-\frac{\mu^2}{\lambda}\Big)^2-\frac{1}{4}F^{a}_{\mu\nu}F^{\mu\nu,a}+\beta^2\Big(1-\sqrt{1+\frac{1}{2\beta^2}B_{\mu\nu}B^{\mu\nu}-\frac{1}{16\beta^4}(B_{\mu\nu}\tilde{B}^{\mu\nu})^2}\Big)
\end{equation}
where:  $A^a_{\mu}$ and $B_{\mu}$ are the $SU(2)$ and $U_Y(1)$ gauge fields respectively; $B^{\mu\nu}$ is the $U_Y(1)$  field strength tensor; $\tilde B_{\mu\nu} = \frac{1}{2}\, \epsilon_{\mu\nu\rho\sigma}\, B^{\rho\sigma}$ (with $\epsilon_{\mu\nu\rho\sigma}$  being the covariant Levi-Civita fully antisymmetric tensor); $F^{\mu\nu,a}$ ($a=1,2,3$) is the $SU(2)$ field strength tensor;  $D_{\mu}=\partial_{\mu}-i\frac{g}{2}\tau^aA^a_{\mu}-i\frac{g'}{2}B_{\mu}$ is the covariant derivative with, $\tau_{a}$, $a=1,2,3,$ the Pauli $2 \times 2$ matrices; $\frac{\tau_{a}}{2}$, $a=1,2,3$ are the SU(2) generators; 
$H$ is the electroweak Higgs doublet. 
The $SU(2)$ and $U_Y(1)$ couplings are given by $g$ and $g'$ respectively, 
with 
\begin{eqnarray}\label{gcoupl}
g^\prime = g \, {\rm tan}\theta_W, 
\end{eqnarray}
where $\theta_W$ is the SM weak mixing angle. 
The Born-Infeld parameter $\beta$ has dimensions of [mass]$^2$. The ESM Lagrangian reduces formally to the SM Lagrangian for $\beta\rightarrow\infty$. 
In the context of microscopic  string theory models, the parameter $\sqrt{\beta} \propto M_s$, the string mass scale. In our phenomenological approach here, we deviate from this restriction, and treat $\beta$ as a free parameter to be constrained by experiment, as we shall discuss in secrtion \ref{sec:energy}. 

We shall be interested in finite energy \emph{classical} monopole solutions of Cho-Maison (CM) type~\cite{CM}, for the Euler-Lagrange equations of ESM for finite $\beta $. In the limit $\beta \to \infty$ one would recover the formal CM monopole solution with \emph{divergent }energy.  
The equations of motion, stemming from (\ref{SMBIL}) read:
\begin{align}
	& D^{\mu}(D_{\mu}H)=\lambda\Big(H^{\dagger}H-\frac{\mu^2}{\lambda}\Big)H,\label{EOMH}\\
	& \partial_{\mu}F^{\mu\nu,a}+g f^{abc}A_{\mu}^{b}F^{\mu\nu,c}=\frac{ig}{2}\Big[H^{\dagger}\tau^{a}(D^{\nu}H )-(D^{\nu}H)^{\dagger}\tau^{a}H\Big], \label{EOMA}\\
	& \partial_{\mu}\Big[\frac{B^{\mu\nu}-\frac{1}{4\beta^2}(B_{\alpha\beta}\tilde{B}^{\alpha\beta})\tilde{B}^{\mu\nu}}{\sqrt{1+\frac{1}{2\beta^2}B_{\alpha\beta}B^{\alpha\beta}-\frac{1}{16\beta^4}(B_{\alpha\beta}\tilde{B}^{\alpha\beta})^2}}\Big]=i\frac{g'}{2}\big[H^{\dagger}(D^{\nu}H)-(D^{\nu}H)^{\dagger}H\big].\label{EOMB}
\end{align}
The following ansatz is used to determine the energy of the charged solutions to this Lagrangian~\cite{CM}:
\begin{align}\label{ansatz}
\notag	& H=\frac{1}{\sqrt{2}}\rho(r)\xi\\
	& A^a_{\mu}=\frac{1}{g}{\cal{A}}(r)\partial_{\mu}t\, \hat \phi^a+\frac{1}{g}(f(r)-1)\, \epsilon_{abc} \, \hat \phi^b(\partial_{\mu}\, \hat \phi^c)\\	& B_{\mu}=-\frac{1}{g'}{\cal{B}}(r)\notag \partial_{\mu}t-\frac{1}{g'}(1-cos(\theta))\partial_{\mu}\psi
\end{align}
where
\begin{align}
\notag	\xi=i\begin{bmatrix}
    sin(\theta/2)e^{-i\psi} \\
    -cos(\theta/2)
\end{bmatrix}.
\end{align}
Here, $r^{a}=(t,r,\theta,\psi)$ are spherical polar coordinates (with  $0 \le \theta  \le \pi $ , $0 \le \phi  < 2\pi $ ) and
\begin{align}
\notag	\hat \phi^a=\xi^{\dagger}\tau^a\xi=-\hat r^a, 
\end{align}
where the circumflex denotes unit vector. 

The ansatz (\ref{ansatz}) is best physically understood if one performs a gauge rotation to the unitary \\ gauge~\cite{CM,CKY}:
\begin{eqnarray}
\xi \quad \rightarrow U \, \xi = \begin{pmatrix} 1 \\ 0 \end{pmatrix}, \quad 
{\rm with} \quad U = \begin{pmatrix} \cos(\theta/2)  &  \sin(\theta/2)\, e^{-i \psi} \\ -\sin(\theta/2)\, e^{i\psi} & \cos(\theta/2) \end{pmatrix}~,
\end{eqnarray}
under which the non-Abelian field transforms to \footnote{We use the vector notation to denote the $SU(2)$ gauge field.} 
\begin{eqnarray}
\vec A_\mu = \frac{1}{g}\,  \begin{pmatrix} -f(r) \, \sin(\psi ) \, \partial_\mu \theta + \sin(\theta)\, \cos(\psi)\, \partial_\mu \psi \\
f(r) \, \cos(\psi ) \, \partial_\mu \theta - \sin(\theta)\, \sin(\psi)\, \partial_\mu \psi \\
{\cal{A}}(r) \, \partial_\mu t  - \big(1 - \cos(\theta) \big)\, \partial_\mu \psi \end{pmatrix}.
\end{eqnarray}
The physical fields, the electromagnetic potential $A_\mu^{\rm em}$ and the neutral $Z^0_\mu$ gauge boson field involve the weak mixing angle $\theta_W$, and are given by~\cite{CM}:
\begin{eqnarray}\label{bwaz}
\begin{pmatrix} A_\mu^{\rm em} \\ Z^0_\mu \end{pmatrix}   = \begin{pmatrix} \cos(\theta_W) & \sin(\theta_W) \\ - \sin(\theta_W) & \cos(\theta_W) \end{pmatrix} \, \begin{pmatrix} B_\mu \\ A_\mu^3 \end{pmatrix} =  \frac{1}{\sqrt{g^2 + (g^\prime)^2}} \, \begin{pmatrix} g & g^\prime \\ - g^\prime & g \end{pmatrix} \, \begin{pmatrix} B_\mu \\ A_\mu^3 \end{pmatrix}, 
\end{eqnarray}
on taking into account (\ref{gcoupl}).
In summary the ansatz (\ref{ansatz}) yields the physical fields of the SM~\cite{CM}:
\begin{eqnarray}\label{phys}
{\rm W_{\mu}^{\pm}}&\equiv&\frac{1}{\sqrt{2}}\Big(A_{\mu}^{1}\mp iA_{\mu}^{2}\Big)=\frac{1\,}{g\,\sqrt{2}}\,\Big(\mp if\left(r\right)\partial_{\mu}\theta\exp\left(\mp i\psi\right)+
\sin(\theta)\,\partial\psi\exp\left(\pm i\psi\right)\Big)\nonumber \\
 A_\mu^{\rm em} &=& e \, \Big(\frac{1}{g^2} \, {\cal{A}}(r) + \frac{1}{(g^\prime)^2}\, {\cal{B}}(r) \Big)\, \partial_\mu t - \frac{1}{e}\Big(1 - \cos(\theta) \Big)\, \partial_\mu \psi, \\
\nonumber Z^0_\mu &=& \frac{e}{g^\prime \, g} \Big({\cal{A}}(r) - {\cal{B}}(r) \Big)\, \partial_\mu t ,
\end{eqnarray}
with 
\begin{equation}\label{elec}
e=g\, \sin(\theta_W) = \frac{g\, g^\prime}{\sqrt{g^2 + (g^\prime)^2}},
\end{equation}
the electron charge.
As can be seen from (\ref{phys}), the spherically symmetric (static) monopole solution of \cite{CM} is characterised by:
\begin{equation}\label{ab0}
{\cal{A}}(r)={\cal{B}}(r)=0. 
\end{equation}
In this case the electromagnetic potential resembles a  Dirac 
point-like monopole, 
\begin{eqnarray}\label{dirac}
A_\mu^{\rm em} = - \frac{1}{e}\Big(1 - \cos(\theta) \Big)\, \partial_\mu \psi,
\end{eqnarray}
but the magnetic charge $g^m$ is \emph{twice} the fundamental Dirac charge; so 
\begin{equation}
g^m = \frac{4\pi}{e} = \frac{4\pi\,\sqrt{g^2 + (g^\prime)^2}}{g\, g^\prime}~,
\label{magch}
\end{equation}
where in the last equality we used (\ref{elec}). 

If (\ref{ab0}) is valid, from (\ref{phys}) the $Z_\mu^0$ configuration vanishes, 
\begin{equation}\label{z0}
Z_\mu^0 =0. 
\end{equation}
and moreover the expressions (\ref{ansatz}) reduce to
\begin{align}\label{anstazmon}
\notag	& H=\frac{1}{\sqrt{2}}\rho(r)\xi\\
& A^a_{\mu}=\frac{1}{g}(f(r)-1)\epsilon_{abc} \, \hat \phi^b(\partial_{\mu} \hat \phi^c)\\
\notag	& B_{\mu}=-\frac{1}{g'}(1-cos(\theta))\partial_{\mu}\psi.
\end{align}
and the equation for the hypercharge gauge boson (\ref{EOMB}) is trivially satisfied.  
\footnote{The right-hand-side (RHS) of (\ref{EOMB}), upon using  (\ref{anstazmon}), is the same as for the CM case \cite{CM,CKY}, and vanishes on account of (\ref{ab0}):
\begin{eqnarray}\label{cmrhs}
{\rm RHS~of}~(\ref{EOMB})= -\frac{(g^\prime)^2}{4}\, \rho(r)^2 \, \Big({\cal{A}}(r)-{\cal{B}}(r) \Big) \quad \stackrel{(\ref{ab0})}{\to} \quad 0~. 
\end{eqnarray}
The monopole solution is characterised by a zero electric field and a spherically symmetric static, radial magnetic field, 
\begin{equation}\label{magfield}
B_r (r) \propto \frac{g^{m}}{r^2}, \quad B_\theta = B_\psi =0~, 
\end{equation}
with $g^m$ the magnetic charge (\ref{magch}). Moreover, given (\ref{z0}), one obtains from (\ref{bwaz}) for the monopole solution: 
\begin{equation}\label{monab}
A(r)_\mu^{\rm em~mono} = \frac{\sqrt{g^2 + (g^\prime)^2}}{g}\, B(r)^{\rm mono}_\mu .
\end{equation}
Equation (\ref{monab}), implies that only the spatial components of $B^{\mu\nu}$ are non-zero and proportional to the magnetic field $B_k$:  $B^{ij} \propto \epsilon^{ijk}\, B_k $, with $\epsilon^{ijk}$ the totally antisymmetric Levi-Civita tensor. Moreover, since the electric field is zero in the monopole case, one has 
\begin{equation}\label{bifields}
B_{\alpha\beta}B^{\alpha\beta} \propto + \vec B(r)^2, \qquad B_{\alpha\beta}\tilde{B}^{\alpha\beta} =0.
\end{equation}
Hence, the left hand side (LHS) of (\ref{EOMB}) involves the derivative of a static function that depends solely on the radial-coordinate $r$; the only potentially non-zero contribution should come when $\mu =r$.  
From this we obtain  for the LHS of (\ref{EOMB}) (retaining only the potentially non-zero terms in the argument of the derivative):
\begin{equation}
\Big({\rm LHS~of}~(\ref{EOMB})\Big)^j \propto \partial_i \, \Big[\frac{\epsilon^{ijk}\, B_k(r)}{\sqrt{1+\frac{1}{2\beta^2}B_{\alpha\beta}B^{\alpha\beta}}}\Big] \quad \stackrel{(\ref{magfield})}{\to} \quad 0,
\end{equation}
since, as already mentioned, the only potentially non-trivial component of the derivative is $i=r$. Thus, equation  (\ref{EOMB}) is trivially satisfied for the monopole solution with (\ref{ab0}) in the Born-Infeld case (and the reader should recall that this is also what happens in the CM case~\cite{CM,CKY}). }
From (\ref{anstazmon}), we also note that the hypercharge-sector  `magnetic field' $\mathcal{B}^Y_i =  \epsilon_{ijk}\,  \partial^j B^k $ corresponding to $B_\mu$, assumes the following for the monopole solution:
\begin{equation}\label{hmf}
\vec{{\mathcal B}^Y} =  \Big(\frac{4\pi}{g^\prime}\Big)^2 \frac{\vec r}{r^3}.
\end{equation}
This has the same singular form (as $r \to 0$) as the monopole magnetic field (\ref{magfield}), but with the magnetic charge being replaced by the `hypermagnetic charge $g_Y^m \equiv \frac{4\pi}{g^\prime}$.
 We shall make use of (\ref{hmf}), when we evaluate the total energy of the Born-Infeld-Cho-Maison-like solution in section \ref{sec:energy}.

From now we will concentrate on the equations of motion for the Higgs field and $SU(2)$ \\ gauge field, (\ref{EOMH}) and (\ref{EOMA}), respectively; these equations coincide with those in the \\ ordinary CM case~\cite{CM}. On using (\ref{anstazmon}), these  become
\begin{subequations}\label{EquationsOfMotion}
\begin{align}
	& {\rho}^{\prime\prime}+\frac{2}{r}{\rho}^\prime-\frac{f^2}{2r^2}\rho=\lambda\Big(\frac{\rho^2}{2}-\frac{\mu^2}{\lambda}\Big)\rho\\
	& {f}^{\prime\prime}-\frac{f^2-1}{r^2}f=\frac{g^2}{4}\, \rho^2 \, f,  \\
	& {\cal{A}}^{\prime\prime} + \frac{2}{r} \, {\cal{A}}^\prime - 2\frac{f^2}{r^2}\, {\cal{A}} = \frac{g^2}{4}\, \rho^2 \, \Big({\cal{A}} - {\cal{B}}\Big)
\end{align}
\end{subequations}
where the prime denotes differentiation with respect to $r$. For the monopole solution, for which (\ref{ab0}) is valid, the third of the above equations is trivially satisfied, yielding zero on both sides. 
The \emph{trivial }solution of the equations of motion (\ref{EquationsOfMotion}), which yields the Dirac monopole, has no \\ W$^\pm_\mu$-bosons, that is
\begin{equation}\label{param}
f(r)=0~, \quad {\rm and} \quad \rho = \rho_0=\sqrt{\frac{2\mu^2}{\lambda}} \ne 0,
\end{equation}
with $\rho_0$ the Higgs field vacuum expectation value (vev) in the broken symmetry phase.

\section{New solutions for Born-Infeld-inspired electroweak dressed magnetic monopoles \label{sec:dressed}}

We shall consider new solutions of (\ref{EquationsOfMotion}) where we still have (\ref{ab0})  but $f(r)$ and $\rho(r)$ are \\ 
allowed to be non-trivial. Such solutions can be interpreted  as Dirac monopoles dressed by  W$^\pm_\mu$-bosons and have not been discussed so far in the literature.\footnote{The case ${\cal{A}}(r) \ne 0,  \rm{and} \  {\cal{B}}(r) \ne 0$ leads to the CM dyon solution~\cite{CM}.} We seek solutions of (\ref{EquationsOfMotion})  for $\rho(r)$ and $f(r)$ which satisfy the following boundary conditions 
\begin{align}\label{bc}	
f(r=0) &=1, \; \;  \rho(r=0)=0,  \nonumber \\
 f(r=\infty) &=0,\; \;  \rho(r=\infty)=\rho_0=\sqrt{\frac{2\mu^2}{\lambda}} \ne 0.
\end{align}
Before further analysis we will rewrite the equations (\ref{EquationsOfMotion})  in terms of dimensionless  quantities:
\[\tilde \rho  = \frac{\rho }{{{\rho _0}}},\]
\[\varepsilon  = \frac{{{g^2}}}{{2\lambda }},\] and
\[x = \mu r.\]
Hence the dimensionless forms of the first two equations of (\ref{EquationsOfMotion}) are\footnote{There is a slight abuse of notation since we still use the notation $f$ and $\tilde \rho$ for functions of $x$.}:
\begin{equation}
\label{e1}
\tilde \rho ''\left( x \right) + \frac{2}{x}\tilde \rho '\left( x \right) - \frac{{{f^2}\left( x \right)}}{{2{x^2}}}\tilde \rho \left( x \right) = \tilde \rho \left( x \right)\left( {{{\tilde \rho }^2}\left( x \right) - 1} \right)
\end{equation}
and
\begin{equation}
\label{e2}
f''\left( x \right) - \frac{{{f^2}\left( x \right) - 1}}{{{x^2}}}f\left( x \right) = \varepsilon {{\tilde \rho }^2}\left( x \right)f\left( x \right)
\end{equation}
where $'$ denotes $\frac{d}{{dx}}$.  The boundary conditions for $\tilde \rho \left( x \right)$ are $\tilde \rho(x=0)=0$ and $\tilde \rho(x=\infty)=1$. It is important to note that, from (\ref{e1}),  $f^{2}\left( x \right)$ is determined in terms of $\tilde \rho $ and its derivatives.

The system of equations (\ref{e1}) and (\ref{e2}) are usually solved numerically~\cite{CM,AK}; however there is a delicate interplay in the small $x$ behaviour of $f\left( x \right)$ and  $\tilde \rho \left( x \right)$ which purely numerical solutions can miss. The equations (\ref{bc}), (\ref{e1}) and (\ref{e2}) represent a boundary value. Unlike initial value problems, boundary value problems are not guaranteed to have a unique solution; in some instances there may be no solution at all. Coupled boundary value problems pose an additional challenge since approximations for  $\tilde \rho$ and $f$ cannot be made independently.
We will first obtain asymptotic expansions as $x \to 0$ and as $x \to \infty $. From these asymptotic expansions and a smooth interpolating function we can evaluate  the energy of the monopole within ESM. 

The matching of the asymptotic expansions for large and small $x$ would result in an interpolating solution; this matching is, however, not straightforward. Consequently we will take a different approach to determining a suitable interpolating function based on a Pad\'e approximant for a small $x$ asymptotic series. We take $\varepsilon \sim .81$ which is obtained from  values of the parameters phenomenologically relevant for SM, namely, $g \simeq 0.65 $
and $\lambda\simeq .26$.

\subsection{Large $x$ asymptotics}
Respecting the boundary conditions (\ref{bc}) we write
$\tilde \rho \left( x \right) = 1 + \tilde \delta \left( x \right)$ with $\left| {\tilde \delta \left( x \right)} \right| \ll 1$. To leading order then (\ref{e2}) becomes
\begin{equation}
\label{e3}
f''\left( x \right) + \frac{{f\left( x \right)}}{{{x^2}}} = \varepsilon f\left( x \right).
\end{equation}
Similarly (\ref{e1}) becomes 
\begin{equation}
\label{e4}
\tilde \delta ''\left( x \right) + \frac{2}{x}\tilde \delta '\left( x \right) - 2\tilde \delta \left( x \right) = \frac{{{f^2}\left( x \right)}}{{2{x^2}}}.
\end{equation}
\subsubsection{Behaviour of $f\left( x \right)$}
The leading  behaviour of (\ref{e3}) is governed by
\begin{equation}
\label{e5 }
f''\left( x \right) = \varepsilon f\left( x \right)
\end{equation}
 and the solution compatible with the boundary conditions is 
$f\left( x \right) = {f_1}\exp \left( { - \sqrt \varepsilon  x} \right).$ In order to include the subleading behaviour we write 
$f\left( x \right) = {f_1}\exp \left( { - \sqrt \varepsilon  x} \right) + \Delta \left( x \right)$ where
\begin{equation}
\label{e6 }
\Delta ''\left( x \right) - \varepsilon {\kern 1pt} \Delta \left( x \right) =  - \frac{{{f_1}}}{{{x^2}}}\exp \left( { - \sqrt \varepsilon  x} \right).
\end{equation}
A particular solution ${\Delta _p}\left( x \right)$ of this inhomogeneous second order differential equation (using the method of variation of parameters) is
\begin{equation}
\label{e7}
{\Delta _p}\left( x \right) = \frac{{\exp \left( { - \sqrt \varepsilon  x} \right)}}{{2\sqrt \varepsilon  }}\int\limits_{}^x {dt\frac{{{f_1}}}{{{t^2}}} \;- \;} \frac{{\exp \left( {\sqrt \varepsilon  x} \right)}}{{2\sqrt \varepsilon  }} \int\limits_{}^x {dt\frac{{{f_1}}}{{{t^2}}}} \exp \left( { - 2\sqrt \varepsilon  t} \right)
\end{equation}
and the resultant $f(x)=f_{\infty}(x)$ is
\begin{equation}
\label{e8}
f_{\infty}\left( x \right) \sim \exp \left( { - \sqrt \varepsilon  x} \right)\left[ {{d_1} - \frac{{{f_1}}}{{2\sqrt \varepsilon  x}}\left( {1 - \frac{1}{{2\sqrt \varepsilon  x}} + O\left( {\frac{1}{{{x^2}}}} \right)} \right)} \right].
\end{equation}

\subsubsection{Behaviour of $\tilde \delta \left( x \right)$}
We shall now consider the inhomogeneous equation (\ref{e4}). In terms of 
\[F\left( x \right) \equiv \frac{{f_\infty ^2\left( x \right)}}{{2{x^2}}}\]
a particular solution ${{\tilde \delta }_p}\left( x \right)$ of (\ref{e4}) is given by
\begin{equation}
\label{e9}
{{\tilde \delta }_p}\left( x \right) = \frac{1}{{2\sqrt 2 }}\left( { - \exp \left( { - \sqrt 2 x} \right)\int\limits_{}^x {t\exp \left( {\sqrt 2 t} \right)F\left( t \right)dt + \exp \left( {\sqrt 2 x} \right)\int\limits_{}^x {t\exp \left( { - \sqrt 2 t} \right)F\left( t \right)dt} } } \right).
\end{equation}
The asymptotic expansion of $\tilde \rho (x)$ is then given by 
$1 + {{\tilde \delta }_p}\left( x \right) + {d_2}\frac{{\exp \left( { - \sqrt 2 x} \right)}}{x}$ where we have added \\ a solution of the homogeneous equation and 
\begin{equation}
\label{e9a}
\widetilde{\delta}_{p}\left(x\right)=-\frac{d_{1}}{2\sqrt{2}x}\left[\frac{f_{1}}{2\sqrt{\varepsilon}}\exp\left(-2\sqrt{\varepsilon}x\right)+\frac{\sqrt{2}f_{1}\left(\sqrt{2\varepsilon}-1\right)+\sqrt{\varepsilon}d_{1}}{2\sqrt{2\varepsilon}\left(\sqrt{2\varepsilon}-1\right)}\exp\left(-2\left(\sqrt{2}-\sqrt{\varepsilon}\right)x\right)\right].
\end{equation}
The dominant exponential in (\ref{e9a})  as $x \to \infty$ is $\exp\left(-2\left(\sqrt{2}-\sqrt{\varepsilon}\right)x\right)$.

\subsection{Small $x$ asymptotics}
We write 
$f\left( x \right) = 1 + {\Delta _0}\left( x \right)$ with 
${\Delta _0}\left( x \right) \to 0$ as $x \to 0$ and the leading behaviour in this limit  is determined by
\begin{equation}
\label{e10}
{\Delta _0}''\left( x \right) - \frac{2}{{{x^2}}}{\Delta _0}\left( x \right) = \varepsilon {\kern 1pt} {{\tilde \rho }^2}\left( x \right).
\end{equation}
For our solution it is important to note that 
${{\tilde \rho }^2}\left( x \right)$ is \emph{not} assumed to be small in comparison with 
${\Delta _0}\left( x \right)$. The remaining equation to be considered is
\begin{equation}
\label{e11}
\tilde \rho ''\left( x \right) + \frac{2}{x}\tilde \rho '\left( x \right) + \left( {1 - \frac{1}{{2{x^2}}}} \right)\tilde \rho \left( x \right) = 0
\end{equation}
which has a solution 
$\tilde \rho \left( x \right) \propto {j_\delta }\left( x \right)\left( { \equiv \sqrt {\frac{\pi }{{2x}}} {J_{\delta  + \frac{1}{2}}}\left( x \right)} \right)$ where 
$\delta  = \frac{{\sqrt 3  - 1}}{2}$. From the power series for Bessel functions we have
\[{j_\delta }\left( x \right) = \frac{{\sqrt \pi  }}{2}{x^\delta }\sum\limits_{m = 0}^\infty  {\frac{{{{\left( { - 1} \right)}^m}}}{{m!}}} \frac{{{{\left( {x/2} \right)}^{2m}}}}{{\Gamma \left( {\delta  + m + 3/2} \right)}}.\]
To low order in $x$
\begin{equation}
\label{e12}
\tilde \rho \left( x \right) \simeq {c_1}{x^\delta }\left( {1 - \frac{{{x^2}}}{{2\left( {2\delta  + 3} \right)}}} \right).
\end{equation}
From (\ref{e10}) and (\ref{e12}) we deduce that 
\begin{equation}
\label{e13}
{\Delta _0}''\left( x \right) - \frac{2}{{{x^2}}}{\Delta _0}\left( x \right) = \varepsilon {\kern 1pt} c_1^2{x^{2\delta }}{\left( {1 - \frac{{{x^2}}}{{2\left( {2\delta  + 3} \right)}}} \right)^2}.
\end{equation}
The relevant particular solution ${\Delta _{0p}}\left( x \right)$ of (\ref{e13}) is \footnote{Since ignoring the right hand side of (\ref{e13}) is not consistent we  need to just consider the particular solution. See the Appendix.}
\begin{equation}
\label{e14}
{\Delta _{0p}}\left( x \right) = \frac{{\varepsilon {\kern 1pt} c_1^2}}{3}{x^{2 + 2\delta }}\left[ {\frac{3}{{2\delta \left( {3 + 2\delta } \right)}} - \frac{{3{x^2}}}{2}{{\left( {15 + 31\delta  + 20{\delta ^2} + 4{\delta ^3}} \right)}^{ - 1}} + O\left( {{x^4}} \right)} \right].
\end{equation}
\subsection{Summary of leading asymptotic solutions}
We shall for convenience gather together the results of our asymptotic analysis.

For small $x$ :

\[\tilde \rho \left( x \right) \sim {c_1}{x^\delta }\left( {1 - \frac{{{x^2}}}{{2\left( {2\delta  + 3} \right)}}} \right),\]
\begin{equation}
\label{e15}
f\left( x \right) \sim 1 + \frac{{\varepsilon {\kern 1pt} c_1^2}}{2}{x^{2 + 2\delta }}\left( {\frac{1}{{\delta \left( {3 + 2\delta } \right)}} - \frac{{{x^2}}}{{15 + 31\delta  + 20{\delta ^2} + 4{\delta ^3}}} + O\left( {{x^4}} \right)} \right).
\end{equation}
It is interesting to note that our asymptotic analysis has revealed  a "bump" in $f\left( x \right)$ for small $x$.

For large $x$:
\begin{equation}
\label{e16}
f\left( x \right) \sim \exp \left( { - \sqrt \varepsilon  x} \right)\left( {{d_1} - \frac{{{f_1}}}{{2\sqrt \varepsilon  x}}\left( {1 - \frac{1}{{2\sqrt \varepsilon  x}} + \frac{1}{{2\varepsilon {x^2}}}} \right)} \right),
\end{equation}
\begin{equation}
\label{e17}
\rho \left( x \right) \sim 1 + {d_2}\frac{{\exp \left( { - \sqrt 2 x} \right)}}{x} - \frac{1}{{2\sqrt 2 }}\left( \begin{array}{l}
\frac{{{d_1}{f_1}}}{{2\sqrt \varepsilon  x}}\exp \left( { - 2\sqrt \varepsilon  x} \right)\\
 + \frac{{\exp \left( { - 2\left( {\sqrt 2  - \sqrt \varepsilon  } \right)x} \right)}}{x}\frac{{{d_1}\left( {\sqrt 2 {f_1}\left( {\sqrt {2\varepsilon }  - 1} \right) + \sqrt \varepsilon  {d_1}} \right)}}{{2\sqrt {2\varepsilon } \left( {\sqrt {2\varepsilon }  - 1} \right)}}
\end{array} \right).
\end{equation}

\subsection{ Higher order small $x$ asymptotic analysis \label{small}}
The structure of the small $x$-behaviour for $f$ and $\rho$ found from the linearised asymptotic analysis for small $x$ in (\ref{e15}) suggests the following ansatz for the nonlinear analysis:
\begin{equation}
\label{e18}
f\left(x\right)=1+\sum_{m=1}^{\infty}\sum_{n=2}^{\infty}a_{mn}x^{2m+n\delta}
\end{equation}
and
\begin{equation}
\label{e19}
\tilde\rho\left(x\right)=\sum_{m=0}^{\infty}\sum_{n=1}^{\infty}b_{mn}x^{2m+n\delta}.
\end{equation}
We  plug in the expressions (\ref{e18}) and (\ref{e19}) into the coupled differential equations (\ref{e1}) and (\ref{e2}) and equate the coefficients for powers of $x$. We shall give algebraic expressions (in terms of $\delta$ ) for some of the coefficients occurring in (\ref{e18}) and (\ref{e19}). However many coefficients become unwieldy and simplify on putting $\delta=0.36602540378(\approx0.4)$. 
The series in (\ref{e18}) and (\ref{e19}) will be truncated at $m=m_{u}=6$ and $n=n_{u}=6$ and so will give a refined small $x$ asymptotic analysis which will form the basis for a Pad\'e style analysis\footnote{The Pad\'e approximation  \cite{GB} consists of converting the formal power series $\sum_{n}c_{n}x^{n}$ to a sequence of rational functions
\be
P_{M}^{N}\left(x\right)=\frac{\sum_{n=0}^{N}A_{n}x^{n}}{\sum_{m=0}^{M}B_{m}x^{m}}.
\ee
The advantage of constructing $P_{M}^{N}\left(x\right)$ is that in many instances $P_{M}^{N}\left(x\right)$ is a convergent sequence as  $N,M \to \infty$ even when  $\sum_{n}c_{n}x^{n}$ is a divergent series.
} 
The coefficients will be given in the Appendix.

Because of the  powers in $y$ ($y \equiv x^{\delta}$) in (\ref{e18}) and (\ref{e19}) a  conventional Pad\'e approximation (PA) is not possible. However since 
$\delta  \simeq \frac{2}{5}$, it is possible to approximate $y$ by $\tilde y \equiv x^{\frac{2}{5}}$. For convenience let us introduce $z \equiv{x^{2}}(=\tilde y^{5}).$ From (\ref{e19}) we can write down the following small $x$-approximation for :

\be \label{fe20}
\begin{gathered}
\tilde  \rho \left( x \right) \simeq {b_{01}} y + z\left( {{b_{11}} y + {b_{13}}{{ y}^3}} \right) + {z^2}\left( {{b_{21}}y + {b_{23}}{y^3} + {b_{25}}{y^5}} \right) \hfill \\
  \quad \quad \quad  + {z^3}\left( {{b_{31}}y + {b_{33}}{y^3} + {b_{35}}{y^5}} \right) + {z^4}\left( {{b_{41}}y + {b_{43}}{y^3} + {b_{45}}{y^5}} \right). \hfill \\ 
\end{gathered} 
\ee
Before we find a PA we will substitute $y$ with  $\tilde y$
 ( and replace $z$ by $\tilde y^{5}$) . The PA will be in the variable $\tilde y$. We shall construct a diagonal PA  in order to be able to satisfy $\tilde \rho(\tilde y=\infty)=1$. It is straightforward to show that the $(5,5)$ PA, $\rho_{P}(\tilde y)$, has the form 
 \be \label{fe21}
 \rho_{P}(\tilde y)=\frac{ \frac{{b_{01}}
   {b_{13}}^2
  { \tilde y}^5}{{b_{11}}^2}-\frac{{b_{01}} {b_{13}}
   {\tilde y}^3}{b_{11}}+{b_{01}}
  {\tilde y}}{-\frac{{b_{11}}
   {\tilde y}^5}{{b_{01}}}+\frac{{b_{13}}^2
   {\tilde y}^4}{{b_{11}}^2}-\frac{{b_{13}}
   {\tilde y}^2}{{b_{11}}}+1}.
\ee
Clearly $ \rho_{P}(\infty)=-\frac{{b_{01}}^2
   {b_{13}}^2}{{b_{11}}^3}$ and by requiring this to be $1$ we obtain $b_{01} \approx 0.725704$. With this value of $b_{01}$ , $\rho_{P}(y)$ will have the correct small $x$ behaviour and the correct constant asymptotic value. However a PA (generically) cannot accurately reproduce the exponential fall off in (\ref{e17}). Since our aim is to find an interpolating solution which correctly reproduces both the leading asymptotic behaviour for small and for large $x$, we will construct the interpolating function $\tilde \rho_{I}$ to be

\be
\label{fe22}
{\rho _I}\left( x \right) = {\rho _P}\left( y \right)\exp \left( { - 2\left[ {\sqrt 2  - \sqrt \varepsilon  } \right]{y^{\frac{1}{\delta }}}\tanh \left( {{y^{\frac{1}{\delta }}}} \right)} \right).
\ee 

Clearly this modifying factor has the correct large $y$ exponential decay and also for  small $y$ does not affect the leading small $y$ behaviour.The Pad\'e approximant should determine the correct behaviour of $\rho$ at intermediate values of $y$.

The corresponding interpolating function $f_{I}(x)$ for $f(x)$ is determined by ($\ref{e1}$):

\be  \label{fe23}
f_I^2\left( x \right) = 2{x^2}\frac{{\rho '{'_I}\left( x \right)}}{{{\rho _I}\left( x \right)}} + 4x\frac{{\rho {'_I}\left( x \right)}}{{{\rho _I}\left( x \right)}} - 2{x^2}\left( {{\rho ^2}_I\left( x \right) - 1} \right).
\ee

\subsection{Interpolating functions\label{InterpolatingFigs}} 

The equations  (\ref{fe22}) and  (\ref{fe23}) are the primary interpolating functions. They are in the form of explicit analytic expressions. However for numerical estimation of the energy of the monopole they are not efficient in terms of computer time. From the plots for the interpolating functions it is clear that the range 
$\left[ {0.1,\,12} \right]$ for $x$
 is sufficient for the asymptotic values to have been essentially reached. It is numerically more efficient to consider a discrete set of points $(x,\tilde \rho(x))$ and $(x, f(x))$ at a spacing of $.1$ in $x$ for the range $\left[ {0.1,\,12} \right]$. We can fit these discrete points with a polynomial and produce interpolating functions which are more efficient for evaluation of the energy of the monopole.

\begin{figure}[t]
 \centering
\includegraphics[clip,width=0.40\textwidth,height=0.15\textheight]{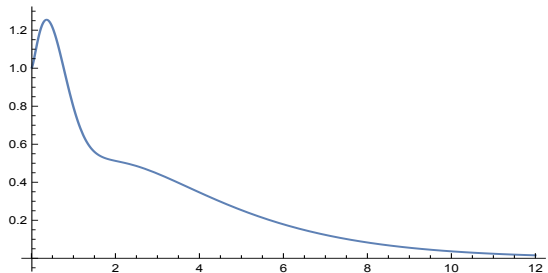} 
\caption{\it Interpolating function for $f(x)$. }\label{fig:Interpolatingf}
\end{figure}

\begin{figure}[t]
 \centering
\includegraphics[clip,width=0.40\textwidth,height=0.15\textheight]{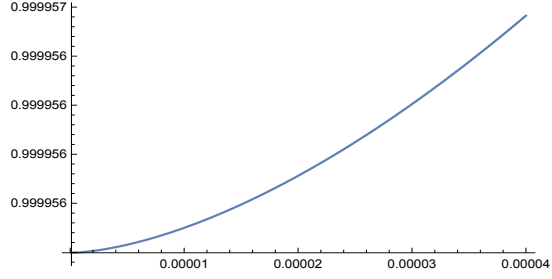} 
\caption{\it Interpolating function for $f(x)$ with $x$ small. }\label{fig:Interpolatingsmallf}
\end{figure}

\begin{figure}[t]
 \centering
\includegraphics[clip,width=0.40\textwidth,height=0.15\textheight]{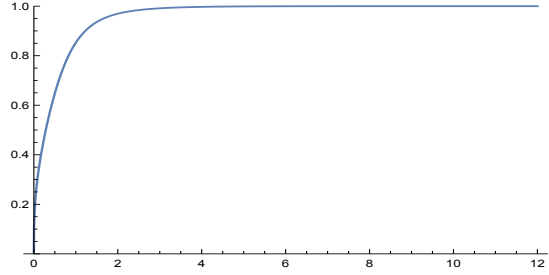} 
\caption{\it Interpolating function for $\tilde \rho(x)$ . }\label{fig:Interpolatingrho}
\end{figure}

We will now give the plots for the primary interpolating functions for  $f\left( x \right)$ and ${\tilde \rho }\left( x \right)$ in Figure \ref{fig:Interpolatingf},  Figure \ref{fig:Interpolatingsmallf} and Figure \ref{fig:Interpolatingrho}.
\medskip

\section{Estimates of the (finite) Monopole Energy \label{sec:energy}}

In this section we estimate the energy of the monopole solution, as this is of importance for phenomenological searches. Form the theory (\ref{SMBIL}), one may evaluate the stress tensor and from this the total energy of the monopole solution. The latter consists of two parts, the first $E_1$ pertains to the kinetic energy of the electromagnetic field (associated with the hypercharge sector) in the non-linear Born-Infeld theory, and the second $E_2$ with the non-Abelian SU(2) gauge and Higgs sectors of the theory. In terms of our parametrisation \eqref{anstazmon}, (\ref{EquationsOfMotion}) and \eqref{param}, 
one has~\cite{AK}
\begin{eqnarray}\label{total}
E_{\rm total}^{\rm mono} &= &E_1 + E_2,  \nonumber \\
E_1& = & 4\pi \, \beta^2 \, \int_0^\infty dr \, \left[\sqrt{r^4+\frac{1}{(g^\prime\, \beta)^2}} - r^2\right] \nonumber \\
E_2&= &4\pi\int_{0}^\infty dr \left(\frac{1}{g^2}\frac{(f^2-1)^2}{2r^2}+\frac{1}{2}\Big(r\,\frac{d\rho}{dr}\Big)^2+\frac{1}{g^2}\, \Big(\frac{df}{dr}\Big)^2+\frac{\lambda r^2}{8}\left(\rho^2-\rho_0^2\right)^2+\frac{1}{4}f^2\rho^2\right),
\end{eqnarray}
where we took into account that in the Born-Infeld hypercharge sector of the monopole solution, only the hypermagnetic field $\vec{{\mathcal B}}$ is non-zero (\ref{hmf}), implying a  Born-Infeld energy $E_1 = \int 4\pi r^2 dr (\beta^2 \sqrt{1 + \frac{({\mathcal B}^Y)^2}{\beta^2} } - \beta^2)$. The result of the integration in $E_1$ can be done analytically by changing integration variable 
$r=\frac{w}{\sqrt{g^\prime\, \beta}}$, ~\cite{initialBI, zwiebach} and using elliptic integrals:
\begin{equation}\label{total2}
E_1  = 4\pi \, \Big(\frac{\beta}{g^{\prime\, 3}}\Big)^{1/2}\,  \int_0^\infty dw \left[ \sqrt{w^4 + 1}-w^2 \right] = 4\pi \, \Big(\frac{\beta}{g^{\prime\, 3}}\Big)^{1/2}\,  \frac{(\Gamma(1/4))^2}{6\, \sqrt{\pi}} \simeq 15.53 \, \Big(\frac{\beta}{g^{\prime\, 3}}\Big)^{1/2}.
\end{equation}
From (\ref{total2}) we thus observe that $E_1$ is finite for any $\beta < \infty$, as a consequence of the non-linearities of the Born-Infeld sector~\cite{zwiebach}. It was this part that produced the infinities in the energy on the Cho-Maison monopole/dyon solutions~\cite{CM}.
Using the SM value $g^\prime = 0.357$, we then obtain
\begin{equation}\label{e1val}
E_1 \simeq 72.81 \sqrt{\beta}. 
\end{equation}
The quantity $E_2$ in (\ref{total}) is also finite, and has been finite also in the CM case~\cite{CM}. 
Using our parametrisation, this quantity can be written as 
\begin{eqnarray}\label{total3}
{E_2} &= & 4\pi \mu \int_0^\infty  {dx\left[ {\frac{{{{\left( {{f^2} - 1} \right)}^2}}}{{2{g^2}{x^2}}} + \frac{{{x^2}}}{\lambda }\tilde \rho {'^{\;2}} + \frac{{f{'^{\;2}}}}{{{g^2}}} + \frac{1}{{2\lambda }}{x^2}{{\left( {{{\tilde \rho }^2} - 1} \right)}^2}}  + \frac{1}{2 \lambda} f^2 \, \tilde \rho^2 \, \right]},
\end{eqnarray}
where the $\prime$ denotes $\frac{d}{dx}$.  By inserting our interpolating solutions into (\ref{total3}), and using the values of the parameters of the Standard Model, $g^\prime/g = {\rm tan}\theta_W~ (\ref{gcoupl}),~g^\prime = 0.357, ~{\rm sin}^2\theta_W =0.2312, ~ \rho_0 \simeq 246.39$~GeV, $\mu = 88.39$~GeV, and $\lambda\simeq .26$, we obtain 
\begin{equation}\label{e2val}
E_2 = 7617~
{\rm GeV},
\end{equation}
which is nearly double (but still of the same order of magnitude as) the value in the CM case~\cite{CM,CKY,AK}. The increase in the value of $E_2$ is a consequence of the difference of our solution  as compared to that of CM, as seen from the figs. \ref{fig:Interpolatingf}, \ref{fig:Interpolatingrho}. 

From (\ref{e1val}), (\ref{e2val}), then, we obtain for the total energy (\ref{total}) of the monopole
\begin{equation}\label{estimate}
E^{\rm mono}_{\rm total} =  (72.81 \, \sqrt{\frac{\beta}{(\rm GeV)^2}} + 7617)~{\rm GeV} 
\end{equation}
In \cite{you2} it was argued that the relatively recent measurements of light-by-light scattering by the ATLAS Collaboration~\cite{hiatl}, exploiting Pb-Pb collisions at LHC, imposes a lower bound on the Born-Infeld parameter of the model (\ref{SMBIL})
\begin{equation}\label{lbb}
\sqrt{\beta} \ge 90 ~{\rm GeV}. 
\end{equation}
From (\ref{estimate}) and (\ref{lbb}), then, we obtain the  following lower bound for the Born-Infeld-Cho-Maison-like monopole mass (=total energy at rest):
\begin{equation}
{\mathcal M}_{\rm mono} \ge  14.17~ {\rm TeV}.
\end{equation}
Since monopoles are produced in pairs with antimonopoles in colliders, on account of (magnetic) charge conservation, our monopole lies out of the detection range of the LHC , but is of potential relevance to future colliders.  

At this stage we stress that $\beta$ is a phenomenological parameter, to be constrained by experiment. However, in the context of microscopic string theory models, though, where the Born-Infeld lagrangian (\ref{SMBIL}) is expected to arise naturally in the low-energy limit, the parameter $\beta \sim M_s^2$, where $M_s$ is the string mass scale. The latter has been constrained by current collider experiments to be at least of ${\mathcal O}(10)$~TeV, thus making the term $E_1$ (\ref{e1val}) dominant over $E_2$ in such a case, leading to a significant increase of the monopole mass $E_{\rm total}^{\rm mono} > 736$~TeV. Such monopoles could be of cosmological relevance, and be potentially detectable in cosmic monopole searches~\cite{mitsou} (for instance, in future cosmic versions of the MoEDAL experiment). If the monopole masses are in the range $9.3 \cdot 10^3 \,\, {\rm TeV} \le M_{\rm mono} \, \le 2.3 \cdot 10^4 \,\, {\rm TeV}$, (with the upper bound associated with 
constraints on the monopole abundance imposed by Big-Bang nucleosynthesis (BBN)) then according to the analysis in \cite{AK}, such cosmic monopoles may have interesting consequences for the early universe, including dynamical generation of matter-antimatter asymmetry.

\section{Conclusions and outlook  \label{sec:concl}} 

In this article, we have discussed some novel semi-analytic (static) monopole solutions in the framework of the phenomenological Lagrangian (\ref{SMBIL}), which constitutes an extension of the SM by a non-linear Born-Infeld Lagrangian for the hypercharge sector only. 
The solutions we found are consistent with asymptotic analysis for the functions $f(x)$ and $\tilde \rho(x)$ characterising the solution
(\ref{anstazmon}), but in contrast to the standard Cho-Maison-like~\cite{CM} solutions in the literature~\cite{AK}, they do exhibit some non-monotonic behaviour for $x$ values near $x=0$. This deviation from the standard numerical solutions though is relatively mild, and does not affect the order of magnitude estimates for the total monopole energy, and the associated phenomenology~\cite{you2}. Nonetheless, from a mathematical point of view our solution is a novel finite-energy monopole solution. Our solutions are analytic, but approximate, as they interpolate between known behaviour for small and large $x$ regions (via appropriate Pad\'e approximants).
Establishing the existence of finite energy monopole solutions is 
important from the experimental point of view, since such solutions can be of relevance for future colliders ( but not the current ones, due to the range of the induced monopole mass which lies outside the capabilities of the LHC ). 

Our analysis in this work should be extended to include dyon solutions, carrying both magnetic and electric charge, following the formalism developed in \cite{CM,AK}. Since the functions $A(r)$ and $B(r)$, characterising the solution (\ref{phys}), are non-zero the analysis is much more involved than in the monopole case. We leave the study of the dyon case for future work. 

Before closing we would like to make an important remark, concerning the finite energy (\ref{total}). In the context of the model (\ref{SMBIL}), the Born-Infeld nature of the hypercharge sector decouples from the $SU(2)$ and Higgs sectors, in the sense that the monopole solution is formally the same as that in the SM case of Cho and Maison~\cite{CM}. It is only the non-linear nature of the Born-Infeld energy $E_1$ that is finite, and proportional to the parameter $\sqrt{\beta}$, which becomes infinite only in the SM limit $\beta \to \infty$. Finite monopole (solitonic) solutions therefore exist for \emph{any value} of $\beta$ in this case. 
However, if one considers effective low-energy field theory models derived from phenomenologically realistic microscopic string theories, then the Born-Infeld non-linear nature is expected to encompass the entire non-Abelian 
gauge group $SU(2) \times U_Y(1)$ and not only the hypercharge $U_Y(1)$. In such cases, the gauge and Higgs sectors mix non-trivially with the hypercharge
and the resulting monopole/dyon solutions are much more complicated than then solutions considered here and in \cite{AK}. Moreover, as discussed in \cite{silva}, in the context of $SU(2)$ Born-Infeld gauge theories, the solitonic monopole/dyon (numerical) solutions exist \emph{only} for values of the Born-Infeld parameter above a critical value, 
$\beta \ge \beta_c$, estimated numerically in \cite{silva}, i.e. the energy diverges for $\beta < \beta_c$.  
Although the analysis in \cite{silva} has been done for simple non-abelian gauge groups $SU(2)$, one expects the above feature to persist in the case where the Born-Infeld sector is extended to include the full standard model non-Abelian group $SU(2) \times U_Y(1)$. At present, such a (non-trivial) extension of the analysis of \cite{silva} is pending. In this respect, however, we should also mention for completeness the model considered in \cite{AK2}, 
according to which two independent Born-Infeld sectors, one for the SU(2) and one for the hypercharge $U_Y(1)$ have been considered, with different parameters $\beta_i$, $i=1,2$. In the analysis of \cite{AK2}, sufficiently large values of the parameter $\beta_2$ for the non-Abelian Born-Infeld sector  have been implicitly assumed, and in this sense the existence of a critical value of $\beta_{2}^c $ cannot be seen.  
Moreover, this is an effective field theory which is however different from the one in a  string theory framework, where the two sectors cannot be separated, and they are both characterised by a common $\beta$. 
We hope to be able to study in detail monopole/dyon solutions in such realistic string-inspired SM extensions, using our semi-analytic methods, in the future.

\section*{Acknowledgments}

We acknowledge discussions with Stephanie Baines.  The authors wish to thank the organisers of the International Conference of New Frontiers in Physics 2018, where results from this work have been presented, for their kind invitation, and for organising such a high-level and stimulating event. This research was funded in part by STFC (UK) research grant ST/P000258/1. N.E.M. also acknowledges a scientific associateship (``\emph{Doctor Vinculado}'') at IFIC-CSIC-Valencia University\ (Spain).

\appendix

\section{Coefficients in small-$x$ asymptotic analysis}

The $b$ coefficients in (\ref{e19}):
\begin{equation} 
\nonumber
b_{11}=-\frac{2b_{01}}{11+10\delta+2\delta^{2}}  
\end{equation}
 \be \nonumber 
b_{12}=0=b_{14}=b_{15}=b_{16}   \ee
\begin{equation} 
 \nonumber 
 b_{13}=\frac{b_{01}^{3}\left(4\delta^{2}+6\delta+\epsilon\right)}{\delta(2\delta+3)\left(18\delta^{2}+30\delta+11\right)}  \end{equation}
\be 
 \nonumber 
b_{21}= \frac{4 b_{01}}{\left(2 \delta ^2+10 
   \delta +11\right) \left(2 \delta
   ^2+18 \delta +39\right)}   \ee
  \be  \nonumber 
b_{22}=0=b_{24}=b_{26}  \ee
 \be  \nonumber 
b_{23}= \frac{2 b_{01}^3 \left(-\frac{\epsilon
   }{\delta  (2 \delta +3) \left(2
   \delta ^2+10 \delta
   +11\right)}-\frac{2 \epsilon
   }{\left(2 \delta ^2+7 \delta
   +5\right) \left(2 \delta ^2+10
   \delta +11\right)}-\frac{2
   \left(\frac{\epsilon }{2 \delta  (2
   \delta +3)}+1\right)}{18 \delta
   ^2+30 \delta +11}-\frac{6}{2 \delta
   ^2+10 \delta +11}\right)}{3 \left(6
   \delta ^2+18 \delta +13\right)}    
   \ee
   \be   \nonumber 
   b_{25}=(0.00625575 + 0.00323926 \epsilon + 0.00292336 \epsilon^2) b_{01}    \ee
\be   \nonumber 
  b_{31}= -\frac{8 b_{01}}{\left(2 \delta ^2+10
   \delta +11\right) \left(2 \delta
   ^2+18 \delta +39\right) \left(2
   \delta ^2+26 \delta +83\right)}      \ee
\be  \nonumber 
b_{32}=0=b_{34}=b_{36} \ee
\be  
 \nonumber 
b_{33}=(0.0015303 + 0.00014715 \epsilon) b_{01}^3 \ee
\be 
 \nonumber 
b_{35}=(-0.00174278-0.00067203\epsilon-0.000413624\epsilon^{2})b_{01}^{5}   \ee
\be 
 \nonumber 
b_{41}=1.61775\times10^{-6} b_{01}    \ee
 \be
 \nonumber 
b_{42}=0=b_{44}=b_{46}   \ee
 \be  
 \nonumber 
 b_{43}=(-0.0000986062 - 4.70038\times10^{-6 }\epsilon) b_{01}^3    \ee
 \be 
 \nonumber 
 b_{45}=(0.000248585+0.00007087280\epsilon+.000030011\varepsilon^{2})b_{01}^{5}    \ee
 \be  \nonumber 
  b_{51}=-1.37892\times10^{-8}b_{01}   \ee
  \be \nonumber
  b_{52}=0=b_{54}=b_{56}    \ee
  \be  \nonumber
  b_{53}=(4.64828\times10^{-6}+1.24034\times10^{-7}\epsilon)b_{01}^{3}   \ee
  \be  \nonumber 
  b_{55}=(-0.0000239827-4.97425\times10^{-6}\epsilon-1.5915\times10^{-6}\epsilon^{2})b_{01}^{5}   \ee
\be  \nonumber 
  b_{61}=8.368\times10^{-11}b_{01}  \ee
 \be  \nonumber 
 b_{62}=0=b_{64}=b_{66}  \ee
 \be  \nonumber 
 b_{63}=-(1.67851\times10^{-7}+2.86179\times10^{-9}\epsilon)b_{01}^{3}   \ee
 \be  \nonumber 
 b_{65}=(1.74488\times10^{-6}+2.5867\times10^{-7}\epsilon+6.95189\times10^{-8}\epsilon^{2})b_{01}^{5} 
 \ee

 We will now list the 'a' coefficients in (\ref{e18}):
 \be  \nonumber 
 a_{12}=\epsilon b_{01}^{2}/(2\delta(3+2\delta))   
 \ee
 \be   \nonumber 
 a_{13}=0 =a_{14} =a_{15}=a_{16}   \ee
 \be  \nonumber 
 a_{22}=-2\epsilon b_{01}^{2}/((5+7\delta+2\delta^{2})(11+10\delta+2\delta^{2})) 
  \ee
 \be  \nonumber 
 a_{23}=0=a_{25}=a_{26} 
  \ee
 \be  \nonumber 
 a_{24}=(0.00732333\epsilon+0.0369758\epsilon^{2})b_{01}^{4} 
 \ee
 \be  \nonumber 
 a_{32}=0.000809972\epsilon b_{01}^{2}   \ee
 \be  \nonumber 
 a_{33}=0=a_{35}=a_{36}  
 \ee
 \be  \nonumber 
 a_{34}=-(0.00115918\epsilon+0.00361289\epsilon^{2})b_{01}^{4} 
  \ee
 \be  \nonumber 
 a_{42}=-0.0000277424\epsilon b_{01}^{2}  \ee
 \be  \nonumber 
 a_{43}=0=a_{45}  \ee
 \be  \nonumber 
 a_{44}=(0.000105727\epsilon+0.000261013\epsilon^{2})b_{01}^{4} 
 \ee
 \be  \nonumber 
 a_{46}=-(0.0000845209\epsilon+0.000354434\epsilon^{2}+0.000438062\epsilon^{3})b_{01}^{6} 
 \ee
\be  \nonumber 
 a_{52}=6.94301\times10^{-7}\epsilon b_{01}^{2}   \ee
 \be \nonumber
 a_{53}=0=a_{55}   \ee
 \be \nonumber
 a_{54}=-(6.89031\times10^{-6}\epsilon+0.0000140244\epsilon^{2})b_{01}^{4} 
 \ee
 \be \nonumber
a_{56}=(0.0000114343\epsilon+0.0000421252\epsilon^{2}+0.0000383327\epsilon^{3})b_{01}^{6} 
\ee
 \be \nonumber
 a_{62}=-1.31163\times10^{-8}\epsilon b_{01}^{2} \ee
 \be \nonumber
 a_{63}=0=a_{65} 
 \ee
 \be \nonumber
 a_{64}=(3.4889\times10^{-7}\epsilon+6.014\times10^{-7}\epsilon^{2})b_{01}^{4} 
  \ee
 \be \nonumber
a_{66}=-(1.0851\times10^{-6}\epsilon+2.57347\times10^{-6}\epsilon^{3}+3.56687\times10^{-6}\epsilon^{2})b_{01}^{6} 
\ee

\end{document}